# Hippocampus mediates conceptual generalization of pain modulation

Dylan Sutterlin-Guindon[1,2], Tor D. Wager*[3], Leonie Koban*[4,5]

Affiliations.

1 Department of Psychology, University of Montreal, Canada

2 Centre de recherche de l'Institut universitaire de gériatrie de Montréal, Canada

3 Department of Psychological and Brain Sciences, Dartmouth College, Hanover, United States

4 Lyon Neuroscience Research Center (CRNL), CNRS, Inserm, Université Claude Bernard, Lyon 1, Bron, France

5 Le Vinatier, Bron, France

* shared senior authorship

Correspondence should be addressed to: Dr. Leonie Koban, SOCIALHEALTH team, Lyon Neuroscience Research Center (CRNL), Centre Hospitalier Le Vinatier, Bâtiment 462 Neurocampus Michel Jouvet, 95 boulevard Pinel, 69500 Bron; email: leonie.koban@cnrs.fr; phone: +33 (0)4 81 10 65 93

## Abstract

Pain is strongly influenced by expectations and learning from previous experience, such as in classical conditioning. Conditioned responses and expectations can generalize to perceptually and conceptually related cues, but how generalization influences pain experience and the neurobiological processing of pain remains unclear. We used fMRI and multilevel mediation analyses to address this question. Thirty-six human participants first learned to associate two visual cues from distinct conceptual categories (e.g., animals vs. vehicles) with high or low levels of heat pain. In a subsequent phase, they were presented novel cues (images, drawings, or words) not previously paired with pain, but which shared the conceptual category of the initial pain-predictive cues. Participants who developed explicit expectations during learning reported greater pain in response to stimuli conceptually related to high- vs. low-pain cues ('generalization stimuli'), demonstrating generalization of cue influences on pain. This effect was mediated by increased pain-related activity to generalization stimuli in the hippocampus, which correlated with individual differences in cue-evoked expectations. A broader network, including areas of the default mode network and striatum, also contributed to conceptual generalization of pain modulation, while threat-related regions such as the amygdala responded to generalization stimuli but did not mediate effects on pain ratings. These findings extend our understanding of expectancy-driven pain modulation by showing how conceptual processes can influence pain and its neurobiological substrates, offering new insight into placebo effects and maladaptive learning in chronic pain.

## Introduction

Pain is the leading reason people seek medical care, imposing substantial costs on health systems and individual well-being (Jackson et al., 2016). However, pain is not merely a response to tissue damage. Modern conceptions of pain suggest that it is an inference constructed in the brain about potential damage (Seymour, 2019), influenced by predictive processes—including placebo and nocebo effects (Büchel et al., 2014; Colloca, 2019; Kaptchuk et al., 2020; Wager & Atlas, 2015)—as well as nociceptive input. Pain, like other experiences and physiological processes related to harm and reward, can be influenced in advance by predictive cues, which is thought to prepare the organism for survival-relevant challenges and opportunities (Eikelboom & Stewart, 1982; Ramsay et al., 2020; Smith et al., 2021). In some cases, cue effects are mediated by expectations, suggesting that conscious, reportable processes are important (Jepma & Wager, 2015; Kirsch, 1985; Koban & Wager, 2016). In other cases, conditioned associations can occur outside of conscious awareness (Benedetti et al., 2003) and be independent of explicit expectations (Goebel et al., 2002; Schafer et al., 2015). This literature demonstrates multiple pathways by which the brain draws on past experience to predict future pain. Identifying the neural mechanisms that support this predictive process is essential for improving treatment and reducing the burden of chronic pain (Motzkin et al., 2023).

Critically, expectations often extend beyond specific prior experiences. Instead, people can generalize across conceptually related situations by relying on abstract similarities (Kumaran, 2012). In aversive learning, for example, fear generalizes to new situations based on their resemblance to previous threats (Dunsmoor & Murphy, 2015). Similar generalization processes may influence pain. People expect pain in novel settings that resemble painful ones (Glogan et al., 2023), and such mechanisms likely contribute to placebo and nocebo effects (Weng et al., 2025). For instance, expectations of relief from one drug may transfer to another medication if the two are perceived as serving the same purpose. Alternatively, if a treatment fails, it may create negative expectations that reduce the effectiveness of future treatments (Zunhammer et al., 2017). Yet, how the brain supports conceptual generalization in pain remains unknown.

Associative learning plays a central role in shaping how we experience pain. In classical conditioning paradigms, a neutral cue used as a conditioned stimulus (CS) is repeatedly paired with either high or low pain. Over time, the CS comes to signal expected pain intensity, and this learned association influences subsequent pain experiences. Cues previously paired with high-intensity pain ($CS_{HIGH}$) compared to low-intensity ($CS_{LOW}$) lead to increased pain ratings, even when the stimulus intensity is kept constant across trials (Atlas & Wager, 2012; Jepma & Wager, 2015; Traxler et al., 2019; Zhang et al., 2019). These learned expectations induced by pain-predictive CSs not only shape subjective pain, but they also change how the brain responds to pain. Functional imaging studies have linked this pain modulation to increased pain-related activity in the thalamus, posterior insula, sensorimotor cortices, and anterior mid-cingulate cortex, as well as in frontal areas involved in attention, value, and expectation (Atlas et al., 2010, 2022; Jepma et al., 2018; Koban et al., 2019; Koyama et al., 2005; Necka et al., 2025). Other systems, including the hippocampus and medial prefrontal cortex, are also key targets for explaining how learned expectations influence pain (Bingel et al., 2022; Motzkin et al., 2021).

While associative learning explains how cues and contexts shape pain processing, it cannot account for how novel cues, never directly paired with pain, can influence pain perception. Generalization allows the predictive value of previously learned stimuli to transfer to new ones (Dunsmoor et al., 2011; Dunsmoor & Murphy, 2015; Honig & Urcuioli, 1981). Most research has focused on perceptual generalization, where generalization stimuli that resemble CSs in their visual or sensory features evoke similar reactions (Kampermann et al., 2021; Lissek et al., 2014; Liu et al., 2019). In contrast, conceptual generalization involves the transfer of learning across abstract or categorical dimensions, independent of perceptual similarity (Kumaran, 2012). Behavioral studies show that both (a) self-reported measures of fear, avoidance and pain expectancy and (b) physiological measures like eye-blink startle and skin conductance responses can generalize based on conceptual similarity to cues (Mertens 2021, Meulders 2017, Meulders & Bennet 2018; Kloos 2022). Also, these self-reported measures and physiological responses can generalize across various modalities, including conceptual categories, physical movements, and real-world representations (Dunsmoor & Murphy, 2015; Glogan et al., 2021; Kloos et al., 2022). In these cases, generalization stimuli never paired with pain can bias subsequent affective responses through expectations developed during initial learning, and these biases can be extinguished if the expected outcome does not follow (Glogan et al., 2019).

Despite these findings, only two studies have directly examined the impact of conceptual generalization on pain perception itself. Koban et al. (2018) found that generalization stimuli from different modalities, such as words or pictures from the same category as a pain-associated cue, could influence pain reports, but only when participants had explicitly learned the initial cue-pain associations (e.g., being aware that $CS_{HIGH}$ predicts more pain as revealed by self-reported expectations). Liu et al. (2019) further showed that these effects consistently altered pain ratings. However, the brain mechanisms underlying this process remain unknown. To address this gap, the present study investigates how the brain mediates the effects of conceptual generalization on pain modulation.

Several candidate brain systems may mediate conceptual generalization effects on pain. On the one hand, generalization effects may rely on regions involved in threat detection and aversive learning, such as the amygdala, insula, anterior mid-cingulate, and striatum (Dunsmoor et al., 2011; Lissek et al., 2014). On the other hand, conceptual generalization effects on pain may depend more on brain regions involved in conceptual and semantic processing, such as the hippocampus and areas of the default mode network (DMN). Neuroimaging studies of perceptual generalization in fear conditioning show that both threat learning-related circuits and regions central for conceptual processing such as the hippocampus and ventromedial prefrontal cortex (vmPFC) contribute to the reinstatement of learned fear responses (Dunsmoor & Murphy, 2015; Onat & Büchel, 2015; Webler et al., 2021, 2024). Across tasks and modalities, the hippocampus has been shown to support cognitive maps—structured representations that capture abstract relationships across spatial, conceptual, and social domains (Constantinescu et al., 2016; Garvert et al., 2017; Sherrill et al., 2023; Tavares et al., 2015). In interaction with key hubs of the DMN, such as the vmPFC, the hippocampus facilitates the integration and generalization of learned associations to novel situations by drawing on high-level features (Bowman & Zeithamova, 2018; Vatansever et al., 2017). This system is thus well-positioned to support the transfer of pain-predictive value from CSs to conceptually related stimuli, and bias pain.

Here, we used fMRI to identify the brain mechanisms underlying pain modulation induced by conceptual generalization. Thirty-six participants first learned to associate visual cues from different conceptual categories (e.g., animals vs. vehicles) with high or low levels of heat pain ($CS_{HIGH}$ and $CS_{LOW}$). In a subsequent generalization phase, they received medium-intensity heat stimulation following novel generalization stimuli (GS) presented in different modalities (images, drawings, or words). Although never paired with pain, these GS belonged to the same conceptual categories as either the high- or low-pain learning cues ($GS_{HIGH\text{-}CAT}$ and $GS_{LOW\text{-}CAT}$). We then used multilevel mediation analyses (Atlas et al., 2010; MacKinnon et al., 2007; Wager et al., 2009) to identify brain regions in which pain-evoked activity was modulated by conceptual generalization and explained trial-by-trial variation in pain ratings. We further tested the hypothesis that individual differences in expectations would moderate these effects, consistent with prior behavioral findings (Kloos et al., 2022; Koban et al., 2018).

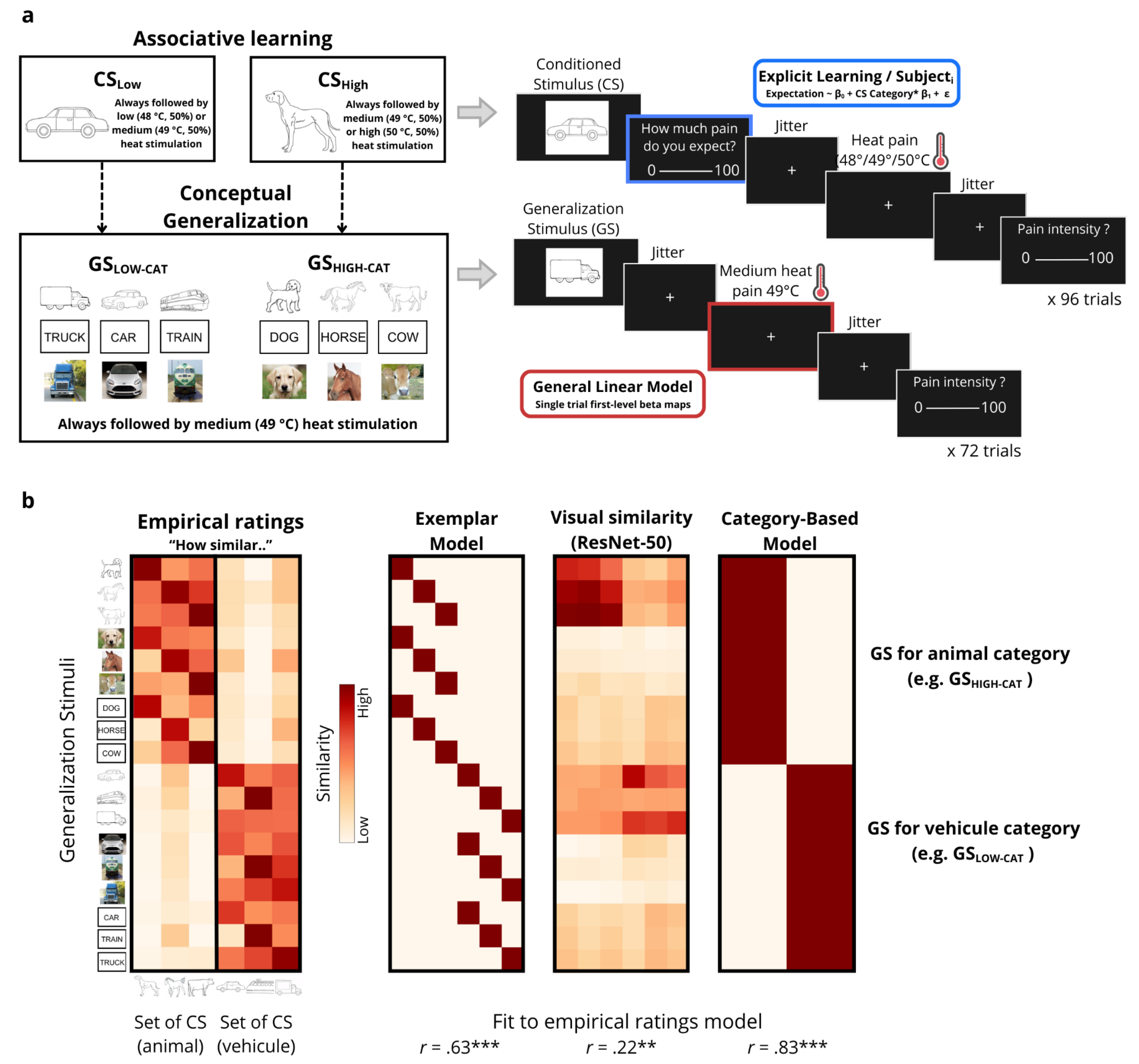

**Figure 1. a)** Paradigm and experimental conditions. *Associative learning task* (previously reported in Koban et al., 2019**).** Participants learned to associate two conditioned stimuli (CS) with different levels of heat pain. The $CS_{LOW}$ cue (e.g., a vehicle) was followed by either low (48 °C, 50% of trials) or medium (49 °C, 50% of trials) heat. The $CS_{HIGH}$ cue was followed by either high (50 °C, 50% of trials) or medium (49 °C, 50% of trials) heat stimulation to the right calf. On each trial, participants first viewed a CS cue (at the same time as social ratings that were independent of the CS cues, see Koban, et al., 2019), rated expected pain (0-100 visual analog scale), received heat stimulation, and then rated pain (0-100 scale). Expectancy ratings from this task (blue box) were used to quantify individual differences in explicit learning of the cue-pain contingency, computed as the slope ($\beta_1$) of expected pain ratings by CS category. *Conceptual generalization task*. In the subsequent generalization phase, participants viewed generalization stimuli—novel exemplars from the same conceptual categories as the CSs—that were never paired with heat stimulation during learning. The $GS_{HIGH\text{-}CAT}$ stimuli (e.g., dog, horse, cow) belonged to the same category as $CS_{HIGH}$, while the $GS_{LOW\text{-}CAT}$ stimuli (e.g., truck, car, train) belonged to the same category as $CS_{LOW}$. Three cue modalities were used during generalization: cartoon, picture, and word. All generalization stimuli were followed by medium heat stimulation (49 °C) without expectation rating. At the end of each trial, participants rated how much pain they experienced, again using a 0-100 VAS). Pain-related brain activity during this phase (red box) was modeled using a single-trial general linear model, producing trial-by-trial beta maps. These maps were then used in subsequent mediation analyses. **b)** Empirical ratings of similarity between conditioned and generalization stimuli compared with three different similarity models. Participants rated the similarity between each of the 18 generalization stimuli and each CS on a continuous scale, without specific instructions about whether to focus on visual or conceptual features. The first heatmap shows the empirical similarity matrix (median ratings across participants), where darker red indicates greater perceived similarity. We tested three models to characterize the structure of these ratings, with Spearman rank correlation, and estimating significance with 10,000 permutations. The *exemplar model* assumed similarity only between identical concepts (e.g., dog CS = dog across all modalities), ignoring category structure. The *ResNet-50 model* was derived from a deep neural network (He et al., 2016) to quantify low-level visual similarity. The category-based model assumed maximal similarity among all stimuli from the same conceptual category (e.g., animals), regardless of exemplar or modality. The *category-based model* best explained the empirical ratings (rho = .83, $p < .0001$), outperforming the exemplar (rho = .63, $p < .0001$) and visual similarity (rho = .21, $p = .01$) models. These results motivated encoding the generalization conditions as a categorical variable ($GS_{LOW\text{-}CAT}$ and $GS_{HIGH\text{-}CAT}$). *** $p < .0001$, ** p = .01

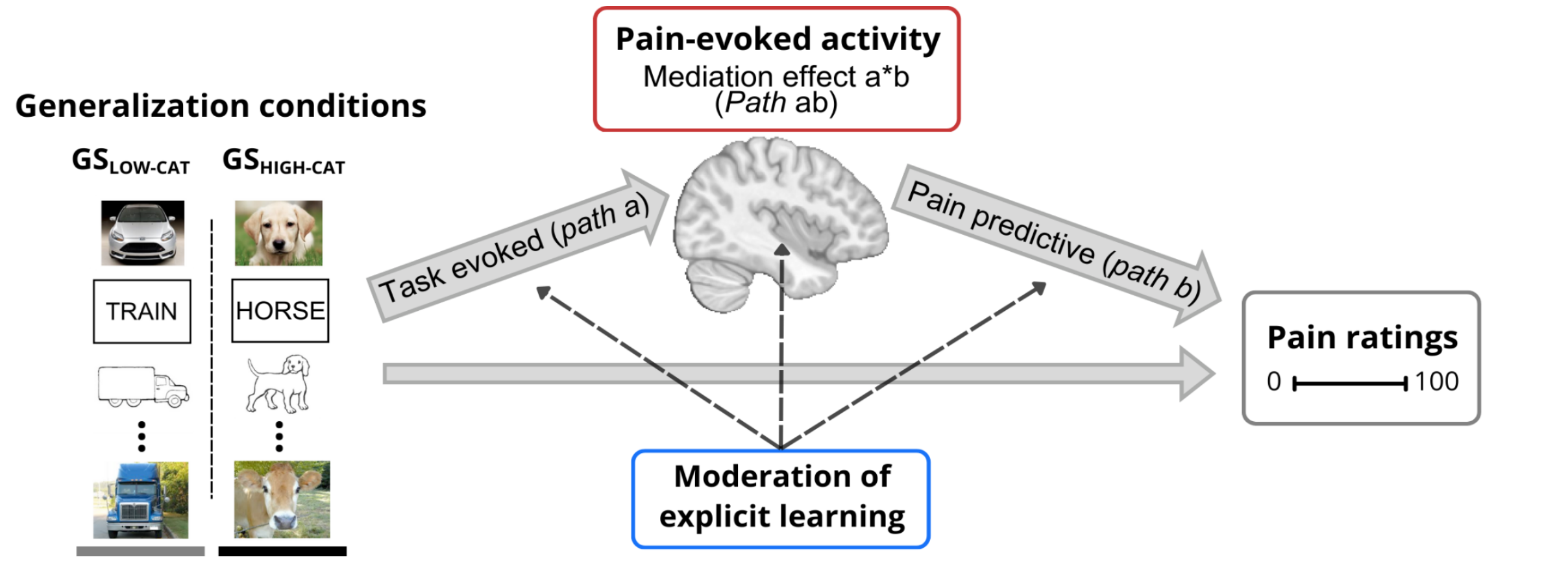

**Figure 2.** Multilevel mass-univariate mediation analysis of pain-evoked brain activity. We conducted a multilevel mass-univariate mediation analysis to identify brain regions involved in conceptual generalization effects on pain processing and pain ratings. Trial-by-trial pain-evoked brain activity was used to test: (1) brain regions showing greater activation during $GS_{HIGH\text{-}CAT} > GS_{LOW\text{-}CAT}$ trials (*path a*), (2) brain activity associated with higher pain ratings on a trial-by-trial basis, controlling for *path a* effects (*path b*), and (3) brain regions formally mediating the effect of conceptual generalization condition on pain ratings (*path ab*). We also tested whether the mediation effect (*path ab*) was moderated by individual differences in explicit learning of the initial CS-pain contingency.

## Results

### *Behavioral results*

During an associative learning procedure, participants learned to associate specific conditioned stimuli (CS) with either high or low levels of pain. To test how these cues influenced pain ratings independently of actual stimulation intensity, medium-intensity (49°C) painful stimuli were presented on 50% of trials within both $CS_{LOW}$ and $CS_{HIGH}$ conditions.

Participants were not told which cues predicted high versus low pain levels. They were only instructed that one cue would predict high and the other cue low pain, and that they had to discover the cue-pain contingency. These instructions led to substantial individual variability in the strength of associative learning. Results from the associative task were previously reported in Koban et al., (2019), but in the present study, the learning strength (degree of explicit learning) was used as a moderator of conceptual generalization. Briefly, on medium-intensity trials, $CS_{HIGH}$ led to higher pain ratings compared to $CS_{LOW}$ ($\beta = 0.42$, 95% CI [0.05–0.79], t(35) = 3.13, $p$ = 0.0363, Cohen's $d$ = 0.52). Participants also reported their pain expectations before each stimulus (see Figure 1a). $CS_{HIGH}$ led to higher expectations than $CS_{LOW}$ on medium-heat trials ($\beta$ = 1.41, 95% CI [0.51–2.31], t(35) = 3.09, $p$ = 0.0038, $d$ = 0.50). These results confirm that pain expectation and pain ratings were successfully altered by the pain-predictive cues (see Figure 3a-b). For a full account of behavioral and neural effects for this task, see Koban et al. (2019).

In the generalization phase, the same participants received medium-intensity pain stimulation following generalization stimuli (GS) that were never paired with pain but belonged to the same conceptual category as either the high- or low-pain CSs. At the group level, conceptual generalization ($GS_{HIGH\text{-}CAT} > GS_{LOW\text{-}CAT}$) did not significantly influence pain ratings ($\beta$ = 0.12 [95% CI -0.35–0.59], t(35) = 0.46, $p$ = .59, $d$ = 0.04). However, consistent with previous studies (Koban et al., 2018), participants who developed more explicit knowledge of the CS-pain contingency (as reflected in higher expectation ratings for $GS_{HIGH\text{-}CAT}$ than $GS_{LOW\text{-}CAT}$) showed larger generalization effects on pain, as revealed by a moderation of generalization effects by individual differences in pain expectation during the associative learning phase ([$GS_{HIGH\text{-}CAT} > GS_{LOW\text{-}CAT}$] x expectancy interaction, $\beta$ = 0.24 [95% CI -0.01–0.49], t(35) = 2.46, $p$ = 0.0293, $d$ = 0.41). These findings indicate that participants who became aware of the CS-pain contingency reported higher pain ratings during $GS_{HIGH\text{-}CAT}$ compared to $GS_{LOW\text{-}CAT}$ trials, showing generalization effects on pain (Figure 3c-d).

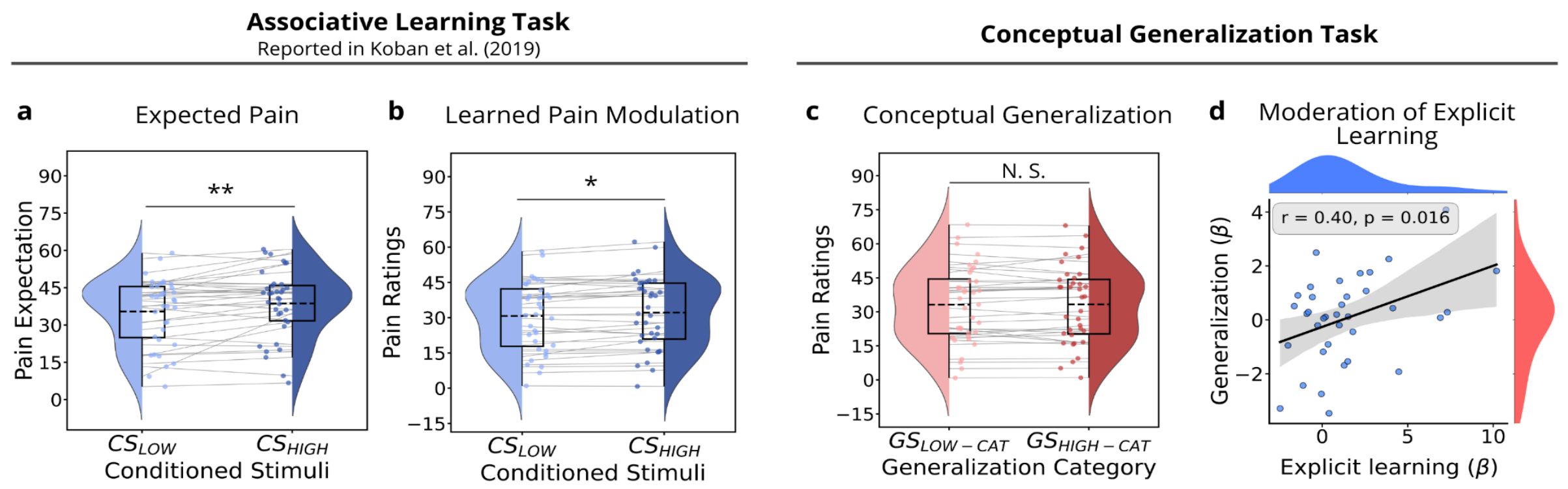

**Figure 3.** We used multilevel general linear models to characterize the behavioral effect of experimental conditions on pain expectancy and pain ratings for the associative learning task (previously reported in Koban et al., 2019) and for the generalization task. **a)** Expected pain during associative learning. Participants reported higher pain expectations for $CS_{HIGH}$ versus $CS_{LOW}$ stimuli during medium-intensity trials (49°C, 50% of trials). Explicit learning scores were quantified as the individual slope ($\beta_1$) of the effect of CS category on expected pain, and used in subsequent moderation analyses. **b)** Learned pain modulation during associative learning. Pain ratings were higher for $CS_{HIGH}$ versus $CS_{LOW}$

stimuli on medium-intensity trials. **c)** Pain ratings during conceptual generalization. Pain ratings did not significantly differ between $GS_{HIGH\text{-}CAT}$ and $GS_{LOW\text{-}CAT}$ conditions overall (n.s.). **d)** Moderation by explicit learning. Individual differences in explicit learning significantly moderated the strength of conceptual generalization effects on pain ratings during GS trials. This moderation is illustrated in the scatterplot by the positive association between individual generalization effect ($\beta$) and individual explicit learning scores ($\beta$). *$p < .05$; **$p < .01$

*Brain mediators of the conceptual generalization effect on pain*

To investigate the brain mechanisms underlying conceptual generalization, we conducted a mediation effect parametric mapping analysis, testing for brain mediation of cue effects on pain within-person and moderation by individual differences in expectation ratings in each brain voxel (Atlas et al., 2010; Wager et al., 2008). The multi-level mediation used trial-by-trial cue values ($GS_{HIGH\text{-}CAT}$ or $GS_{LOW\text{-}CAT}$) as the initial variable, trial-by-trial brain activity as the mediator, and trial-by-trial pain reports as the outcome at the first (within-person) level of the model. *Path a* reflects cue effects on the brain, *Path b* brain effects on pain, and the mediation effect *ab* reflects their joint strength. The second level tests the significance of each path (*a*, *b*, and *ab*) with participant as a random effect, along with moderation of each path by individual differences in explicit learning (average difference in pain expectation for $CS_{HIGH}$ - $CS_{LOW}$) as a second-level moderator (i.e., moderated mediation; see Figure 2). We report significant brain regions for each path (*a*, *b*, and *ab*) FDR corrected at $q < .05$. Full statistical results, including peak coordinates, Z-values, and cluster sizes, are provided in Supplementary Tables 1-3.

*Generalization stimuli effect on pain-evoked brain activity (Path a)*

*Path a* tested whether brain activity during pain differed as a function of generalization conditions ($GS_{HIGH\text{-}CAT} > GS_{LOW\text{-}CAT}$). This contrast revealed strongest activations ($q < 0.05$ FDR corrected) in the bilateral amygdala and in the left mid-posterior insula (Figure 4a). Additional regions showing increased activation included the right premotor cortex, right primary motor cortex, left posterior hippocampus, and subgenual anterior cingulate cortex (sgACC). These findings indicate that $GS_{HIGH\text{-}CAT}$ stimuli (i.e., conceptually related to $CS_{HIGH}$) elicited stronger pain-related brain responses in these regions. In contrast, a small cluster in the right dorsolateral prefrontal cortex (dlPFC) showed a negative effect, indicating reduced activation during $GS_{HIGH\text{-}CAT} > GS_{LOW\text{-}CAT}$ trials.

*Pain-associated brain activity (path b)*

*Path b* tested which brain regions are associated with trial-by-trial variations in subjective pain ratings, controlling for stimulus type ($GS_{HIGH\text{-}CAT} > GS_{LOW\text{-}CAT}$). As expected, pain-predictive brain activity was observed in regions classically associated with pain processing ($q < 0.05$ FDR corrected, Figure 4b), including positive associations in the bilateral anterior and posterior insula, anterior midcingulate cortex (aMCC), thalamus, primary and secondary somatosensory cortices, motor areas, brainstem, and cerebellum. Additional positive effects were observed in the left dlPFC, bilateral visual cortices, and, to a lesser extent, the bilateral superior parietal cortices and left temporal pole. These findings confirm that increased activation in pain-related and higher-order regions was related to higher subjective pain ratings, independent of stimulus type. No significant negative associations were found.

*Brain mediators of conceptual generalization effects on pain ratings (path ab)*

*Path ab* identifies brain regions that formally mediated the relationship between generalization stimuli category and pain ratings. Significant positive mediation effects ($q < 0.05$ FDR corrected) were found across a distributed network, including areas associated with semantic and conceptual processing, memory and valuation. Strongest effects were found in the bilateral ventral striatum (including the nucleus accumbens and caudate), thalamus, left hippocampus and left temporopolar cortex. Additional effects were found in frontal regions, including the lateral and anterior frontal poles and supplementary motor area. Moreover, regions within the default mode network (Alves et al., 2019) also showed significant positive mediation effects, notably the left retrosplenial cortex, bilateral middle temporal gyri, and bilateral posterior parietal cortex. Finally, mediation effects were observed in the precuneus and cerebellum, particularly in crus I, crus II, and lobule IV.

In addition to the mediation effect observed in the left hippocampus, this area also showed a significant, positive moderation effect by explicit learning, as indicated by a positive correlation between the person-level *ab* effect and individual differences in pain expectation ($CS_{HIGH}$ - $CS_{LOW}$) during the associative learning task (Figure 4d). Thus, participants that had explicitly learned the prior CS-pain associations showed stronger hippocampal mediation effects of cue category on pain.

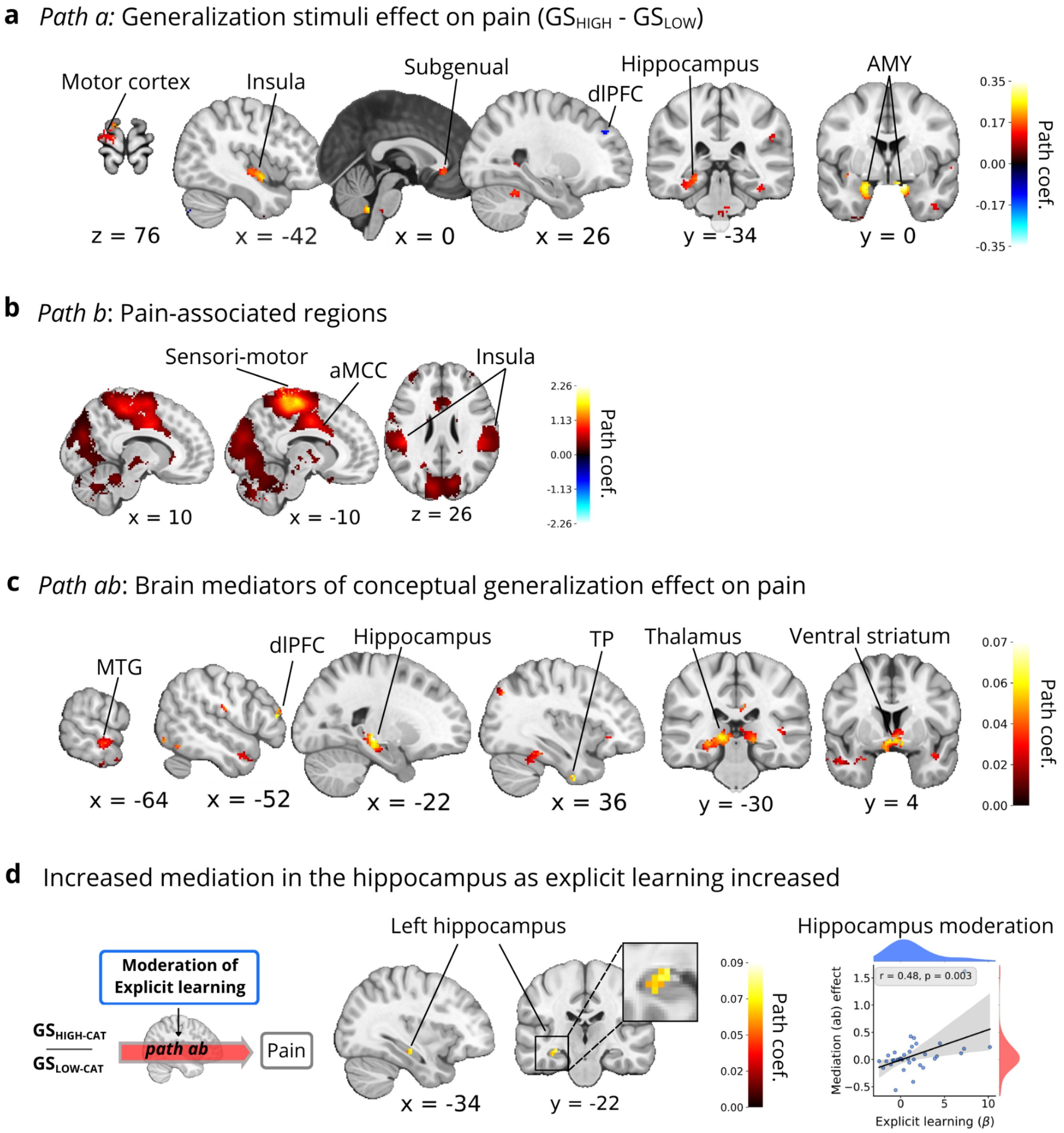

**Figure 4.** Multilevel mass-univariate mediation analysis of conceptual generalization of pain. All results are displayed at an FDR-corrected threshold of $q < .05$. **a)** Effect of generalization condition on pain-related brain activity (*path a*). Regions showing significantly greater activity during pain for $GS_{HIGH\text{-}CAT}$ compared to $GS_{LOW\text{-}CAT}$ trials included the bilateral amygdala, left mid-posterior insula, subgenual anterior cingulate cortex (sgACC), left posterior hippocampus, and right motor cortices. A small cluster in the right dorsolateral prefrontal cortex (dlPFC) showed the opposite pattern, with greater activity for $GS_{LOW\text{-}CAT}$. **b)** Pain-associated brain activity (*path b*). Regions where trial-by-trial activation during pain was positively associated with subjective pain ratings (controlling for generalization condition) included the bilateral anterior and posterior insula, anterior midcingulate cortex (aMCC), thalamus, primary and secondary somatosensory cortices, motor areas, brainstem, cerebellum, left dlPFC, bilateral visual areas and bilateral superior parietal cortex. **c)** Brain mediators of conceptual generalization effects on pain (*path ab*). Regions where activation formally mediated the effect of generalization condition on pain ratings included bilateral ventral striatum (including nucleus accumbens and caudate), thalamus, left hippocampus, left temporopolar cortex, lateral and anterior frontal poles, supplementary motor cortex, bilateral mid-temporal gyri, and bilateral posterior parietal

cortex. Activation in these regions for $GS_{HIGH-CAT} > GS_{LOW-CAT}$ trials was associated with increased pain ratings. **d)** Moderation of hippocampal mediation by explicit learning. The strength of hippocampal mediation (*path ab*) was positively correlated with individual differences in explicit learning of CS-pain associations. This suggests that hippocampal mediation of conceptual generalization effects on pain are stronger for individuals who have developed explicit expectations about the pain-predictive value of the CSs. The scatterplot illustrates this positive relationship for visualization purposes only. Each dot represents a subject.

## Discussion

Pain is shaped by prior expectations and predictions, often formed through learned associations between events and painful experiences. These expectations can generalize beyond the original learning setting, biasing how pain is perceived and processed in the brain. Conceptual generalization—defined as the transfer of learned associations based on abstract or high-level similarity—offers a framework for understanding this process. Clarifying how conceptual generalization shapes pain, both behaviorally and neurally, is important, as it may contribute to maladaptive responses such as nocebo effects and to chronic pain (Kampermann et al., 2021; Meulders, 2019; Weng et al., 2025).

In this study, we examined whether novel stimuli that share a conceptual resemblance to pain-predictive cues, but were never paired with any level of pain, could modulate pain perception, and used fMRI to probe the brain mechanisms underlying this effect. Although these generalization stimuli (GS) did not affect self-reported pain ratings at the group level, participants who had explicitly learned the original stimulus-pain associations reported greater pain in response to GS from the same conceptual category as high-pain learning stimuli. These behavioral effects were mirrored in the brain. Pain-evoked activity in the hippocampus was higher for generalization stimuli of the same conceptual category as the high-pain learning cue and mediated the conceptual generalization effect on pain ratings. This effect was stronger among participants who had developed more explicit expectations. Additional mediation effects were observed in brain areas related to valuation, such as the ventral striatum, and in areas related to semantic processing, especially regions of the default mode network. Our findings thereby identify a potential mechanism by which expectations driven by conceptual relationships can influence pain.

### *Conceptual generalization relies on expectations*

Our work builds on previous behavioral studies (Koban et al., 2018; Liu et al., 2019) showing that altered pain reports to novel stimuli are not automatic. Instead, they depend on contingency awareness, whether people have explicitly learned the initial stimulus-pain associations during conditioning. This suggests that conceptual generalization reflects higher-order inference rather than passive or implicit processes (Dunsmoor & Murphy, 2015; Kumaran, 2012). We further show that both behavioral and brain responses to generalization stimuli scale with the strength of prior learning. Participants who formed stronger stimulus-pain expectations during conditioning exhibited greater generalization effects and stronger hippocampal modulation during pain.

Fear learning studies show that contingency awareness moderates conditioning (Baeuchl et al., 2019; Klucken et al., 2009; Mertens, Boddez, et al., 2021). Our findings extend this to conceptual generalization, suggesting that generalization across contexts depends on explicit knowledge of cue-outcome relationships. This mechanism aligns with research showing that pain-related expectations generalize across various settings (Glogan et al., 2023; Kloos et al., 2022; Mertens et al., 2021; Meulders et al., 2017; Meulders & Bennett, 2018). Our results support a cognitive mechanism grounded in abstract representations and explicit expectations that shapes pain processing.

*Hippocampus supports pain generalization through conceptual processes*

Mediation analyses identified the left mid-to-posterior hippocampus as a key region linking conceptual generalization to pain modulation. It was more activated during pain following a generalization stimulus conceptually related to $CS_{HIGH}$ (*path a*), and mediated the effects of stimulus category on increased pain reports (*path ab*), especially for participants who had developed explicit CS-pain expectations during learning. While anterior hippocampus encodes predictive value of pain-associated cues (Wimmer & Büchel, 2021) and contextual pain features (Caston et al., 2023), the mid-posterior segment supports associative processing, memory retrieval, and perceptual integration (Kim et al., 2015; Robinson et al., 2015). Our results align with models of the hippocampus as a hub for cognitive maps, in which abstract relational representations (e.g., category membership) support inference in novel situations (Constantinescu et al., 2016; Garvert et al., 2017; Sherrill et al., 2023). Garvert et al. (2017) implicated the mid-posterior hippocampus in such inferential processes, consistent with our peak mediation and moderation effects.

In predictive processing models of pain, prior expectations shape incoming nociceptive signals, amplifying or attenuating pain based on learned associations and expected value (Seymour, 2019). Our findings support the idea that the hippocampus contributes to this predictive coding-based pain modulation, by linking cues to knowledge-based conceptual maps. Such processes could help explain how pain-related expectations persist, generalize, and influence new experiences.

This mechanism has implications for chronic pain. The fear-avoidance model (Vlaeyen, 2015; Vlaeyen & Linton, 2000) describes how fear of pain leads to avoidance behavior, which can generalize to similar contexts or movements. This generalization of pain-related behavior and cognition can block extinction of maladaptive behaviors and recovery, ultimately contributing to the maintenance of pain (see Meulders, 2019, for a review). Our findings suggest that the hippocampus may support this process by enabling conceptual generalization of learned expectations.

Supporting this, animal models have identified the hippocampus as a critical site in the development of chronic pain. Blocking neurogenesis in this region prevents the transition to chronic pain after nerve injury (Apkarian et al., 2016; Mutso et al., 2012, 2014). In humans, altered hippocampal-prefrontal connectivity has been linked to pain chronification (Vachon-Presseau et al., 2016). More recently, the functional connectivity of the posterior hippocampus with other posterior regions predicted chronic pain development after an injury in individuals without a prior history of chronic pain, potentially reflecting altered affective learning and expectations in these individuals (Branco et al., 2024). Together, these findings suggest that the hippocampus may be a key site where pain regulation, learning processes and conceptual mechanisms interact.

Our paradigm was specifically designed to isolate these conceptual influences on pain. Generalization stimuli were never directly paired with pain and were presented in three distinct modalities (images, drawings, and written words), minimizing perceptual overlap (see Figure 1b). Their only shared feature was a conceptual category (e.g., animal or vehicle) with the initial pain-predictive conditioned stimuli. Brain activity was measured during pain *itself*, not during cue presentation or anticipation, ensuring that hippocampal activation reflected the abstract pain value inferred from conceptual similarity to prior learning. This supports the role of the hippocampus in top-down modulation of nociception via memory-based expectations, likely in interaction with other memory, valuation, and semantic systems.

*A dual mechanism supporting conceptual generalization*

Beyond the hippocampus, our mediation analysis identified a broader network involved in conceptual generalization, including the ventral striatum, thalamus, temporal pole, and default mode network (DMN). DMN regions, involved in semantic integration and self-referential processing, likely support top-down inferences linking conceptual knowledge to sensory input (Koban et al., 2021; Menon, 2023). Their recruitment aligns with evidence that the DMN enables context-specific decision-making and inference based on learned rules (Bowman & Zeithamova, 2018; Vatansever et al., 2017; Zeithamova & Bowman, 2020). DMN-hippocampal interactions also support memory retrieval and recollection of prior experiences (Raichle, 2015; Spreng et al., 2010).

The ventral striatum also emerged as a significant mediator, suggesting that valuation and motivational systems contribute to conceptual generalization of pain modulation. While typically linked to reward learning, the striatum has also been implicated in aversive learning (Klein et al., 2022) and in pain valuation (Baliki et al., 2010, 2013). Mediation in the hippocampus and striatum partially overlapped with 'pro-pain' regions in the SIIPS1 model of endogenous pain modulation, but also showed overlap with 'anti-pain' regions (Woo et al., 2017). Hippocampal-striatal interactions support generalization and inference in reward learning (Colas et al., 2022; Garvert et al., 2023; Wimmer & Shohamy, 2012). Our results suggest that conceptual representations in the hippocampus interact with affective value via striatal circuits during generalization. These results also raise the possibility that hippocampal-striatal interactions in pain generalization reflect the influence of explicit expectations, consistent with previous findings of striatal modulation by explicit conceptual influences (such as instructions) in reinforcement learning and decision-making (Koban et al., 2017; Li et al., 2011; Staudinger & Büchel, 2013).

In contrast, regions involved in threat detection and somatosensory processing, such as the amygdala, mid-posterior insula, and motor cortex, showed increased activation to generalization stimuli conceptually related to $CS_{HIGH}$ compared to $CS_{LOW}$. However, these regions did not mediate the effect of stimulus category on pain ratings or interact with individual differences in explicit learning. Their activity likely reflects generalized threat or salience responses rather than expectation-based modulation, consistent with findings that the amygdala responds more to conditioned than instructed cues (Atlas et al., 2016; Koban et al., 2019), and supports perceptual threat generalization (Lissek et al., 2014; Webler et al., 2021).

Together, these findings point to a dual-mechanism account of conceptual generalization in pain. A low-level, aversive-sensory circuit (amygdala, insula, motor cortex) encodes generalized high-pain representations that may remain unconscious, while in parallel, a higher-order conceptual network (hippocampus, striatum and DMN) retrieves abstract category-based expectations that directly shape subjective pain through top-down modulation.

*Implications for nocebo hyperalgesia*

Our findings suggest a novel mechanism by which nocebo effects, i.e., increased pain driven by expectations, may generalize to new cued contexts. Previous studies have shown that the hippocampus interacts with pain-related networks when more pain is expected, and that this activity correlates with individual differences in pain expectations (Bingel et al., 2011, 2022; Jensen et al., 2015; Kong et al., 2008). These effects are typically observed in response to direct conditioning, and they tend to generalize most strongly across perceptually similar cues or environments (Kampermann et al., 2021; Weng et al., 2025). We extend this by showing that pain responses generalize across conceptually related, but perceptually distinct stimuli, even without direct pain pairing. Our results mirror nocebo mechanisms, where expectations amplify pain without changes in nociceptive input. This hippocampal-driven generalization may explain why people experience pain in novel settings symbolically linked to prior pain.

In addition to hippocampal contributions, other regions involved in self-referential and social cognition, particularly within the DMN, may support other forms of expectancy-based pain modulation. DMN regions have been implicated in integrating socially and conceptually acquired pain-related expectations, such as through observational learning (Schenk & Colloca, 2020) and interpersonal interactions (Ellingsen et al., 2023). Our findings may inform broader research on how high-level influences shape pain experience.

*Future directions*

Future work could examine conceptual generalization in chronic pain populations, where greater generalization of pain and pain-related fear is expected (Meulders et al., 2014, 2017). We expect stronger DMN mediation in these individuals, consistent with evidence of heightened self-referential processing (Fiúza-Fernandes et al., 2025).

We used expectation ratings during learning but not during generalization, to avoid biasing pain reports. Future studies could include occasional ratings to better assess expectations and allow computational modeling of conceptual generalization.

Finally, we note that the generalization categories in our study were restricted to animal and vehicle categories. Testing broader and more ecologically valid categories, such as real-life actions (Glogan et al., 2023) or social cues (Koban et al., 2019), would be beneficial to clarify how conceptual generalization shapes pain in everyday contexts.

*Conclusion*

This study shows that stimulus-pain associations can generalize based on abstract conceptual relationships and bias how pain is experienced. Novel stimuli that were conceptually, but not perceptually, related to pain-predictive cues increased pain ratings, especially in participants who had developed more explicit expectations during initial learning. This behavioral effect was mirrored in the hippocampus, where pain-evoked activity reflected conceptual generalization, predicted pain ratings, and scaled with expectation strength. A broader network related to valuation and conceptual processing also contributed, while threat-related regions responded to conceptual generalization stimuli but were not linked to subjective pain ratings. Together, these findings highlight how conceptual associations can shape pain in novel contexts through expectations, with implications for understanding the generalization of nocebo effects and maladaptive pain responses.

## Methods

*Participants*

A total of 38 healthy volunteers, with no previous psychiatric, neurological, or pain disorders, took part in the study. Two participants were excluded from the analyses due to high movement artifacts or technical complications during the experiment. The final sample comprised 36 individuals (20 females and 16 males, mean age = 27.1 years, age range 18–50 years). All participants provided informed consent and received compensation for their time. The study adhered to the Declaration of Helsinki and was approved by the Institutional Review Board at the Department of Psychology and Neuroscience, University of Colorado Boulder.

*Stimuli*

The learning stimuli (conditioned stimuli, or CSs) consisted of three cartoon images of animals and three of vehicles. For each participant, only one animal and one vehicle image were shown during the learning phase, with one of them assigned to the high-pain condition ($CS_{HIGH}$) and the other to the low-pain condition ($CS_{LOW}$). The two learning stimuli and the assignment to high versus low pain was completely counterbalanced across participants.

The generalization stimuli (GS) comprised 18 novel cues that were not shown during the learning phase. These included three animals (cow, horse, dog) and three vehicles (car, train, truck). Each concept was represented in three different visual display modalities (drawings, photographs, and written word), resulting in nine GSs per category. GS were categorized based on their conceptual relationship to the CS assigned to high or low pain. The set of generalization stimuli conceptually related to the category of the $CS_{HIGH}$ was labeled $GS_{HIGH\text{-}CAT}$, and the ones conceptually related to the category of the $CS_{LOW}$ were labeled $GS_{LOW\text{-}CAT}$.

*Materials and procedure*

Participants first completed a classical conditioning task designed to induce pain modulation through learned associations, by pairing visual stimuli (cartoon images of animals or vehicles) with different levels of noxious heat (see Figure 1a: *Associative Learning*). One stimulus ($CS_{LOW}$) was paired with low to medium heat (48°C or 49°C), and the other ($CS_{HIGH}$) with medium to high heat (49°C or 50°C). Conditions ($CS_{HIGH}$ versus $CS_{LOW}$; e.g. , animal versus vehicle) were counterbalanced across participants and selected from sets of three cartoons to reduce image-specific effects (see *stimuli*). To isolate the effect of learned pain modulation, medium-intensity heat (49°C) was used on half the trials with each CS. This design allowed comparison of pain responses to identical medium intensity heat stimulation as a function of CSs. On each trial, participants first rated their expected pain on a visual analog scale (VAS; 0-100), anchored at "no pain at all" to "worst pain". After receiving the heat stimulus, they rated their experience of pain using the same scale. The learning task included 96 trials (for details, see Koban, et al., 2019).

In the generalization phase, participants were shown novel visual stimuli that were conceptually related to the CSs but had not been shown previously. Half of the 18 GS were conceptually related to each participant's $CS_{HIGH}$, and half to their $CS_{LOW}$ (see Figure 1a: *Conceptual generalization task*). Each GS was presented for 3 seconds, followed by a jittered interval (2–4 s). They were then followed by a medium-intensity heat stimulus (49°C), delivered on the right leg, using a 27-mm diameter fMRI-compatible CHEPS thermode monitored by a Medoc Pathway system (Medoc, Israel). For each trial, after the heat stimulus, participants rated their pain using the same VAS as in the learning phase. The generalization task included 72 trials, divided into four scanning runs of 18 trials each. The thermode location was changed between each run to prevent sensitization.

At the end of the experiment, participants completed a similarity rating task involving all visual stimuli used in the experiment, including CSs that were not part of their own learning phase (see *stimuli*). They were asked to rate how similar each GS was to each CS, using a VAS ranging from "not similar at all" to "completely similar". No mention was made of visual versus conceptual similarity in the instructions to avoid biasing the assessment. These ratings were used to validate that GSs and CSs from the same conceptual category (e.g., animals or vehicles) were perceived as similar (see *conceptual similarity* below). Due to a technical error, eleven participants only rated half of the potential combinations between GSs and CSs. Therefore the group average similarity matrix was used for further analyses.

*fMRI acquisition and preprocessing*

Functional brain activity was recorded using a Siemens TrioTim 3T scanner, capturing the brain in 26 interleaved transverse slices with 3.4-mm isotropic voxels. Images were acquired with a T2* weighted EPI GRAPPA sequence (TR = 1.3 s, TE = 25 ms, flip angle = 50°, FOV = 220 mm) and preprocessed using SPM8, following a standard procedure that included motion correction, slice-timing correction, spatial normalization to MNI space, and spatial smoothing with an 8-mm FWHM Gaussian kernel. Preprocessed images were resampled to 3x3x3 mm voxel size. For spatial normalization, T1-structural MPRAGE images (1 mm isotropic voxels) were first coregistered to the mean functional image and then normalized to the SPM template using unified segmentation.

*Statistical analyses*

We used multilevel general linear models (GLMs) to examine how experimental conditions influenced pain ratings. All models included participant-specific random intercepts and slopes to account for individual differences in baseline pain sensitivity and in learning rates. Code for all analyses is available at https://github.com/canlab.

*Learning task*

To test whether the learning task resulted in modulation of pain, we focused on trials with medium-intensity heat stimulation. This design allowed a direct comparison of pain ratings for identical stimuli as a function of stimulus type ($CS_{HIGH}$ versus $CS_{LOW}$, coded as 1 and -1). A significant difference between $CS_{HIGH}$ versus $CS_{LOW}$ trials indicated that pain ratings were influenced by the learned pain predictive value of each cue.

We also tested whether stimulus type ($CS_{HIGH}$ versus $CS_{LOW}$) influenced pain expectation ratings for medium-intensity trials. The multilevel GLM estimated a random slope for each participant, reflecting how strongly expected pain was influenced by the stimulus type. These participant-level slopes (beta estimates) were used as a between-person measure of explicit learning and were used as moderators for the analysis of behavior and fMRI data in the subsequent generalization task, in line with the results from a series of three previous behavioral studies (Koban et al., 2018) that have shown that explicit learning plays a key role in the generalization of pain modulation.

*Generalization task*

The goal of the conceptual generalization task was to test whether pain ratings were modulated by the category of the generalization stimulus, i.e., whether cues conceptually related to $CS_{HIGH}$ or $CS_{LOW}$ modulated pain responses to identical noxious stimuli. We used multilevel GLMs to estimate the effect of the generalization stimulus category ($GS_{HIGH\text{-}CAT}$, $GS_{LOW\text{-}CAT}$, coded as 1 and –1) on pain ratings. Explicit learning scores from the associative task were included as a second-level moderator to test whether individual learning differences modulated the generalization effect.

*Conceptual similarity*

To characterize the structure of perceived similarity between generalization and conditioned stimuli, we analyzed participants' similarity ratings collected at the end of the experiment (see *Materials and procedure*). Participants rated how similar each GS was to each CS, including CSs not seen during their own learning phase. No reference was made to conceptual characteristics or visual features in the instructions.

To reduce the influence of missing data caused by a software malfunction, we computed an empirical similarity matrix using the median rating, instead of the mean, across participants for each GS–CS pair (see Figure 1b). Visual inspection of this matrix suggested that participants grouped stimuli based on conceptual categories, and not modality. For example, the concept of "dog," "horse," and "cow," regardless of modality (word, picture, or cartoon), were rated as highly similar to animal CSs, but dissimilar to vehicle stimuli.

To formalize this observation and to characterize the structure of these empirical ratings, we compared the similarity matrix to three models using Spearman correlations (see Figure 1b). The exemplar model was a binary matrix assuming maximal similarity (1) only between CS-GS pairs that referred to the same object (e.g., dog word, dog picture, dog cartoon), and no similarity (0) to all other items, even if they belonged to the same conceptual category (e.g., animal). The category model also used binary coding, but assumed maximal similarity between all items from the same conceptual category (e.g., all animals), regardless of modality or exemplar. Finally, the ResNet-50 visual model estimated similarity based on low-level image features, using cosine similarity between image embeddings of each CS and GS. ResNet-50 is a deep convolutional neural network designed for image classification and feature extraction (He et al., 2016; https://huggingface.co/microsoft/resnet-50). This network was chosen over other models because its architecture has been shown to simulate early and mid-level visual processing in the human brain (Schrimpf et al., 2018).

Among the three, the category model best explained the empirical ratings ($r = .83$, $p < .0001$), outperforming both the exemplar model ($r = .63$, $p < .0001$) and the ResNet-50 model ($r = .21$, $p < .0001$). These results confirm that participants organized stimuli based on conceptual category rather than perceptual similarity or only based on object/examplar specificity. This supports our use of the $GS_{HIGH\text{-}CAT}$ and $GS_{LOW\text{-}CAT}$ categorization in both behavioral and brain analyses.

*fMRI analyses*

We used SPM12 to estimate trial-by-trial pain-evoked brain activity using a single-event general linear model (GLM). Six motion parameters were included as covariates. This approach produced one beta map for each of the 72 generalization trials. Although single-trial models provide greater sensitivity to trial-level variation than block designs, they are more susceptible to noise. To address this, we calculated the variance inflation factor (VIF) for each trial to assess multicollinearity with regressors. Trials with a VIF greater than 4 were excluded, resulting in the removal of 33 beta maps (out of 2592 total). The remaining beta maps served as input for the multilevel mediation analyses.

We applied voxel-wise multilevel mediation analysis using the CANlab Mediation Toolbox (https://github.com/canlab/MediationToolbox) to identify brain regions linking generalization conditions ($GS_{LOW\text{-}CAT}$ and $GS_{HIGH\text{-}CAT}$) to changes in pain rating. This approach models participants as random effects and tests whether a mediator (M) statistically explains the relationship between an independent variable (X) and an outcome (Y) (Atlas et al., 2010; MacKinnon et al., 2007). In our model, the GS category (coded as -1 for $GS_{LOW\text{-}CAT}$ and 1 for GS$_{\text{HIGH-CAT}}$) served as the independent variable (X), pain ratings as the outcome (Y), and pain-evoked beta maps as the mediator (M). As in behavioral analyses, explicit learning scores were included as a second-level moderator to test if mediation was influenced by individual differences in expectation during learning (see Figure 2).

The mediation model estimates three paths, each with a distinct interpretation. *Path a* tests whether the GS category influences brain activity. This path identifies voxels with significant differences in activation between generalization trials (GS$_{\text{HIGH-CAT}}$ > GS$_{\text{LOW-CAT}}$) and corresponds to a standard GLM contrast and reflects pain-related responses modulated by conceptual generalization. Positive or negative coefficients indicate whether a given region shows greater or lesser activation during GS$_{\text{HIGH-CAT}}$ trials. *Path b* tests whether trial-by-trial variation in brain activity is associated with subjective pain ratings, controlling for GS category. This identifies pain-predictive regions, regardless of the generalization condition. *Path ab,* the mediation path, tests whether changes in activity in a given voxel account for the effect of GS category on pain ratings. This identifies brain regions where activation during GS$_{\text{HIGH-CAT}}$ versus GS$_{\text{LOW-CAT}}$ trials explains increased pain ratings. Because this is a multilevel model, it captures both within and between-subject effects. For example, positive mediation coefficients indicate that participants who show stronger activation for GS$_{\text{HIGH-CAT}}$ also tend to report higher pain ratings on those trials.

*Moderation*

We included participants' explicit learning scores as a second-level moderator in the mediation model. The moderation tested whether individual differences in learning strength influenced the magnitude of mediation effects across the brain. We examined positive moderation effects for *path ab*, identifying regions where the mediation was stronger among participants who developed explicit expectations about the pain-predictive values of CSs during the learning phase (see *Learning task*).

*Significance*

Significant clusters in each path were identified using False Discovery Rate (FDR) correction at $q < 0.05$. Across all analyses, the voxel-wise mediation models tested mediation effects in 586,836 voxels. For this number of comparisons, an FDR of $q < 0.05$ corresponded to $p < .003$. We report all significant clusters with a minimum size of $k > 3$.

## Contributions

L.K. and T.D.W. conceptualized the experiment. L.K. acquired the data. D.S.G. and L.K. analyzed the data. T.D.W. and L.K. obtained funding and supervised the project. D.S.G. and L.K. wrote the original draft and created the figures. All authors contributed to the interpretation of the data and to the editing of the paper.

## Conflicts of interest

The authors declare no conflict of interests.

**Acknowledgements**

This study was funded by the National Institutes of Health (NIH; R37MH076136, PI: TDW) and an ERC Starting Grant SOCIALCRAVING (101041087, PI: LK). DSG was supported by Unifying Neuroscience and Artificial Intelligence and by Fonds de recherche du Québec—Nature et technologies excellence grants, as well as by Mitacs Globalinks fellowship programs. Views and opinions expressed are however those of the authors only and do not necessarily reflect those of the European Union or the European Research Council. Neither the European Union nor the granting authority can be held responsible for them. The funders had no role in study design, data analysis, manuscript preparation, or publication decisions.

**Supplementary information:**

# Hippocampus mediates conceptual generalization of pain modulation

Dylan Sutterlin-Guindon, Tor D. Wager*, Leonie Koban*

*equal contribution

Contact: Leonie.Koban@cnrs.fr

**Supplementary Table 1.** Path a effects from mediation analyses (FDR-corrected, q < 0.05). This table reports positive and negative effects of the $GS_{HIGH-CAT}$ > $GS_{LOW-CAT}$ contrast during medium-intensity heat stimulation. Brain regions are labeled using the combined anatomical atlas developed by the CANlab group, which integrates several parcellation schemes. Cortical regions (Ctx) are based on 468 parcels from the multimodal Glasser et al. (2016) atlas, basal ganglia subdivisions from Pauli et al. (2016, 2018), cerebellar regions from Diedrichsen et al. (2009), brainstem subdivisions from Fairbust et al. (2007) and from Shen et al. (2013). The full combined anatomical atlas can be downloaded from https://github.com/canlab/Neuroimaging_Pattern_Masks/tree/master/Atlases_and_parcellations/2018_Wager_combined_atlas, with additional documentation available at https://sites.google.com/dartmouth.edu/canlab-brainpatterns/brain-atlases-and-parcellations/2018-combined-atlas. In addition, large-scale functional networks are labeled based on the atlas by Schaefer et al. (2018).

| Region | Volume (mm3) | X | Y | Z | maxZ | Large-scale region network |
|---|---|---|---|---|---|---|
| ***Path a* positive effects ($GS_{HIGH-CAT}$ > $GS_{LOW-CAT}$)** | | | | | | |
| AMY L | 3592 | -22 | -4 | -20 | 3.94 | Amygdala |
| AMY R | 992 | 14 | -2 | -22 | 3.94 | Amygdala |
| NAC R | 128 | 6 | 2 | -16 | 3.21 | Basal ganglia |
| Bstem Ponscd | 144 | 8 | -36 | -38 | 3.23 | Brainstem |
| Bstem Pons R | 144 | 12 | -20 | -36 | 3.3 | Brainstem |
| Bstem Midb L | 280 | -8 | -24 | -18 | 3.45 | Brainstem |
| Bstem Pons R | 160 | 12 | -16 | -20 | 3.02 | Brainstem |
| Cblm IX R | 824 | 4 | -44 | -40 | 3.54 | Cerebellum |
| Cblm VI R | 1496 | 32 | -44 | -26 | 3.9 | Cerebellum |
| Ctx a24 R | 600 | 0 | 28 | -4 | 3.74 | Cortex Default Mode A |
| Ctx PHA3 L | 400 | -34 | -34 | -18 | 4.05 | Cortex Default Mode C |
| Ctx PHA1 L | 568 | -26 | -32 | -12 | 3.46 | Cortex Default Mode C |
| Ctx 6ma L | 224 | -16 | -6 | 78 | 3.16 | Cortex Fronto Parietal A |
| Ctx TGv L | 272 | -28 | 2 | -48 | 3.89 | Cortex Limbic |
| Ctx TGd R | 1160 | 44 | 4 | -38 | 4.11 | Cortex Limbic |
| Ctx TGd R | 128 | 54 | 6 | -38 | 3.52 | Cortex Limbic |
| Ctx TF R | 424 | 40 | -28 | -20 | 3.42 | Cortex Limbic |
| Ctx 6d L | 1888 | -24 | -16 | 76 | 3.78 | Cortex Somatomotor A |
| Ctx A5 R | 160 | 58 | -6 | -2 | 3.07 | Cortex Temporal Parietal |
| Ctx PoI2 L | 1656 | -40 | -10 | -6 | 4.19 | Cortex Ventral Attention A |
| Ctx PF R | 448 | 50 | -30 | 30 | 3.6 | Cortex Ventral Attention A |
| ***Path a*** **negative effects ($GS_{HIGH-CAT}$ > $GS_{LOW-CAT}$)** | | | | | | |
| Ctx 9 46d R | 192 | 26 | 44 | 36 | -2.96 | Cortex Ventral Attention B |

**Supplementary Table 2.** Path b effects of mediation analyses (FDR corrected $q < 0.05$). Positive and negative effects of brain activity associated with pain ratings on a trial-by-trial basis, while controlling for generalization conditions ($GS_{HIGH-CAT}$ and $GS_{LOW-CAT}$).

| Region | Volume (mm3) | X | Y | Z | maxZ | Large-scale region or network |
|---|---|---|---|---|---|---|
| ***Path b* positive effects (increased activity predicts more pain, controlling for *path a*)** | | | | | | |
| Multiple regions | 75024 | 46 | -10 | 6 | 5.19 | Basal ganglia |
| Multiple regions | 83688 | -42 | -10 | 10 | 4.97 | Basal ganglia |
| Cau R | 2520 | 14 | 26 | 2 | 3.89 | Basal ganglia |
| Caudate Cp R | 368 | 18 | -4 | 24 | 3.46 | Basal ganglia |
| Bstem Pons R | 272 | 8 | -18 | -24 | 3.28 | Brainstem |
| Bstem Midb L | 280 | -8 | -24 | -16 | 3.51 | Brainstem |
| Bstem Midbd R | 1408 | 14 | -22 | -6 | 3.86 | Brainstem |
| Cblm VIIIa R | 1128 | 28 | -62 | -48 | 3.75 | Cerebellum |
| Cblm CrusII L | 472 | -32 | -66 | -46 | 4.18 | Cerebellum |
| Cblm VIIIb L | 280 | -22 | -46 | -44 | 3.23 | Cerebellum |
| Cblm V L | 3216 | -22 | -36 | -34 | 4.24 | Cerebellum |
| Cblm CrusI R | 1208 | 36 | -74 | -26 | 3.8 | Cerebellum |
| Cblm CrusI R | 160 | 26 | -90 | -26 | 3.17 | Cerebellum |
| Ctx TGd L | 744 | -48 | 4 | -38 | 4.38 | Cortex Default Mode B |
| Ctx TGd L | 160 | -38 | 16 | -40 | 3.43 | Cortex Default Mode B |
| Ctx TGd L | 152 | -36 | 14 | -36 | 3.23 | Cortex Default Mode B |
| Ctx PGi L | 2352 | -44 | -52 | 8 | 3.94 | Cortex Default Mode B |
| Ctx MIP R | 208 | 20 | -58 | 36 | 3.3 | Cortex Dorsal Attention A |
| Ctx IP2 L | 792 | -30 | -56 | 34 | 3.82 | Cortex Fronto Parietal A |
| Ctx 111 R | 464 | 22 | 36 | -10 | 4.01 | Cortex Fronto Parietal B |
| Ctx PeEc R | 504 | 18 | 0 | -32 | 3.57 | Cortex Limbic |
| Ctx OFC L | 392 | -8 | 16 | -26 | 3.42 | Cortex Limbic |
| Multiple regions | 117240 | 2 | -24 | 58 | 7.17 | Cortex Somatomotor A |
| Ctx 4 R | 6504 | 46 | -10 | 42 | 4.37 | Cortex Somatomotor A |
| Ctx RI L | 608 | -42 | -40 | 14 | 3.41 | Cortex Somatomotor B |
| Ctx 9 46d L | 5776 | -34 | 56 | 18 | 3.97 | Cortex Ventral Attention B |
| Ctx a9 46v R | 2672 | 38 | 48 | 18 | 3.68 | Cortex Ventral Attention B |
| Ctx V2 R | 616 | 26 | -94 | -16 | 3.75 | Cortex Visual Central |
| Ctx LO1 R | 872 | 36 | -92 | 4 | 3.41 | Cortex Visual Central |
| Ctx V2 R | 176 | 18 | -94 | 12 | 3.31 | Cortex Visual Central |
| Multiple regions | 144832 | -4 | -72 | -2 | 5.3 | Cortex Visual Peripheral |
| Ctx DVT L | 192 | -22 | -68 | 20 | 2.98 | Cortex Visual Peripheral |
| Thal VL | 392 | -14 | -16 | 0 | 3.31 | Diencephalon |
| Thal VA | 144 | -12 | -6 | 6 | 3.03 | Diencephalon |
| Thal Intralam | 1176 | 2 | -22 | 12 | 4.04 | Diencephalon |
| No label | 136 | -48 | -46 | -6 | 3.68 | No description |
| No label | 288 | -32 | -44 | 6 | 3.71 | No description |
| No label | 224 | -30 | -50 | 14 | 3.1 | No description |

| **Path b negative effects (reduced activity is associated with more pain, controlling for path a)** | | | | | | |
|---|---|---|---|---|---|---|
| Ctx OFC L | 128 | -8 | 32 | -30 | -3 | Cortex Limbic |

**Supplementary Table 3.** *Path ab* effects of mediation analyses (FDR corrected $q < 0.05$). Positive mediation effects indicate that increased pain-evoked activity during generalization ($GS_{HIGH-CAT} > GS_{LOW-CAT}$) was associated with more pain ratings.

| Region | Volume (mm3) | X | Y | Z | maxZ | Large-scale region or network |
|---|---|---|---|---|---|---|
| ***Path ab* positive effects (mediation of conceptual generalization effect on pain ratings)** | | | | | | |
| Multiple regions | 9376 | 2 | 4 | -6 | 5.64 | Basal ganglia |
| V Striatum R | 624 | 16 | 22 | -4 | 4.71 | Basal ganglia |
| Cau R | 616 | 24 | -32 | 14 | 5.03 | Basal ganglia |
| Caudate Ca R | 160 | 20 | 20 | 6 | 3.91 | Basal ganglia |
| Caudate Cp L | 496 | -8 | 12 | 16 | 6.35 | Basal ganglia |
| Caudate Cp L | 248 | -12 | 18 | 14 | 4.17 | Basal ganglia |
| Bstem Med R | 320 | 2 | -38 | -46 | 3.62 | Brainstem |
| Bstem Ponscv R | 464 | 2 | -24 | -46 | 4.06 | Brainstem |
| Multiple regions | 11488 | -14 | -30 | -6 | 8.04 | Brainstem |
| Cblm CrusII R | 864 | 14 | -84 | -40 | 4.26 | Cerebellum |
| Cblm CrusI R | 264 | 44 | -50 | -36 | 3.75 | Cerebellum |
| Cblm CrusII R | 744 | 10 | -88 | -32 | 5.15 | Cerebellum |
| Cblm CrusI R | 408 | 44 | -78 | -30 | 3.97 | Cerebellum |
| Cblm VI R | 1808 | 34 | -46 | -24 | 5.04 | Cerebellum |
| Cblm CrusI L | 736 | -44 | -40 | -28 | 4.03 | Cerebellum |
| Cblm CrusII R | 568 | 16 | -90 | -26 | 4.06 | Cerebellum |
| Cblm I IV R | 208 | 10 | -34 | -14 | 3.59 | Cerebellum |
| Ctx TE1a R | 2008 | 60 | -14 | -16 | 5.1 | Cortex Default Mode A |
| Ctx 10d L | 192 | -20 | 66 | 14 | 3.04 | Cortex Default Mode A |
| Ctx 31pd R | 200 | 10 | -50 | 26 | 3.24 | Cortex Default Mode A |
| Ctx RSC R | 672 | 6 | -32 | 26 | 4.22 | Cortex Default Mode A |
| Ctx 7m L | 1464 | -8 | -56 | 30 | 3.56 | Cortex Default Mode A |
| Ctx PGs R | 2200 | 40 | -70 | 38 | 4.61 | Cortex Default Mode A |
| Ctx 8Ad L | 248 | -20 | 32 | 36 | 3.82 | Cortex Default Mode A |
| Ctx 8Ad R | 392 | 26 | 22 | 40 | 4.63 | Cortex Default Mode A |
| Ctx PGs L | 1064 | -44 | -68 | 42 | 4.15 | Cortex Default Mode A |
| Ctx 8Ad R | 272 | 14 | 34 | 42 | 3.96 | Cortex Default Mode A |
| Ctx TGd L | 6384 | -40 | 8 | -28 | 7.92 | Corte Default Mode B |
| Ctx TE1a_L | 800 | -64 | -16 | -28 | 5.29 | Cortex Default Mode B |
| Ctx 47s L | 128 | -30 | 18 | -26 | 3.33 | Cortex Default Mode B |
| Ctx 45 L | 568 | -58 | 30 | 16 | 3.9 | Cortex Default Mode B |
| Ctx POS1 R | 192 | 8 | -52 | 14 | 3.06 | Cortex Default Mode C |
| Ctx PGi R | 160 | 40 | -58 | 28 | 3.36 | Cortex Default Mode C |
| Ctx RSC L | 824 | -6 | -16 | 30 | 5.62 | Cortex Default Mode C |
| Ctx TE2p L | 344 | -48 | -24 | -26 | 3.87 | Cortex Dorsal Attention A |
| Ctx FFC L | 400 | -50 | -60 | -16 | 3.73 | Cortex Dorsal Attention A |
| Ctx TE1m L | 176 | -54 | -28 | -18 | 3.9 | Cortex Fronto Parietal B |
| Ctx a10p R | 328 | 22 | 62 | -10 | 3.65 | Cortex Fronto Parietal B |
| Ctx a47r R | 576 | 30 | 62 | 2 | 3.96 | Cortex Fronto Parietal B |
| Ctx POS2 R | 440 | 0 | -70 | 26 | 3.2 | Cortex Fronto Parietal C |
| Ctx TGv R | 1016 | 42 | -6 | -42 | 4.01 | Cortex Limbic |
| Ctx TGd R | 472 | 24 | 16 | -40 | 4.88 | Cortex Limbic |
| Ctx TE2a L | 184 | -52 | -26 | -28 | 3.42 | Cortex Limbic |
| Ctx TGd R | 528 | 42 | 6 | -22 | 5.68 | Cortex Limbic |
| Ctx 13l L | 288 | -32 | 32 | -20 | 3.88 | Cortex Limbic |
| Ctx TF R | 192 | 38 | -26 | -14 | 3.37 | Cortex Limbic |
| Ctx 10pp R | 640 | 18 | 56 | -10 | 4.17 | Cortex Limbic |
| Ctx 3a L | 960 | -50 | -14 | 24 | 4.56 | Cortex Somatomotor A |
| Ctx 4 L | 152 | -50 | -6 | 24 | 3.33 | Cortex Somatomotor A |
| Ctx 1 L | 992 | -58 | -24 | 50 | 4.14 | Cortex Somatomotor A |
| Ctx 3b R | 152 | 46 | -14 | 42 | 3.48 | Cortex Somatomotor A |
| Ctx 6mp L | 536 | -16 | -22 | 64 | 3.83 | Cortex Somatomotor A |
| Ctx 6v L | 280 | -64 | 6 | 28 | 5.18 | Cortex Somatomotor B |
| Ctx STGa R | 432 | 52 | 8 | -22 | 3.32 | Cortex Temporal Parietal |
| Ctx A5 L | 2480 | -64 | -16 | -8 | 6.98 | Cortex Temporal Parietal |
| Ctx A5 L | 200 | -62 | 0 | -10 | 4.98 | Cortex Temporal Parietal |
| Ctx STSdp R | 1144 | 44 | -28 | 4 | 4.08 | Cortex Temporal Parietal |
| Ctx AAIC R | 1992 | 26 | 14 | -18 | 3.8 | Cortex Ventral Attention A |
| Ctx 6r L | 320 | -62 | 12 | 20 | 4.59 | Cortex Ventral Attention A |
| Ctx SCEF L | 472 | -4 | -2 | 50 | 3.63 | Cortex Ventral Attention A |
| Ctx SCEF R | 392 | 4 | 8 | 52 | 3.8 | Cortex Ventral Attention A |
| Ctx AVI L | 320 | -30 | 18 | -8 | 3.51 | Cortex Ventral Attention B |
| Ctx AVI L | 864 | -30 | 26 | -8 | 3.93 | Cortex Ventral Attention B |
| Ctx FOP5 R | 384 | 38 | 30 | -4 | 3.71 | Cortex Ventral Attention B |
| Ctx 46 L | 568 | -50 | 40 | 16 | 3.55 | Cortex Ventral Attention B |
| Ctx 9 46d L | 488 | -28 | 38 | 22 | 3.95 | Cortex Ventral Attention B |
| Ctx 46 R | 280 | 42 | 50 | 26 | 3.65 | Cortex Ventral Attention B |
| Ctx PIT L | 560 | -52 | -72 | -10 | 3.71 | Cortex Visual Central |
| Ctx V1 L | 1048 | -4 | -58 | 4 | 3.75 | Cortex Visual Peripheral |
| Ctx ProS L | 552 | -28 | -54 | 4 | 3.47 | Cortex Visual Peripheral |
| Ctx V6A R | 320 | 20 | -86 | 42 | 3.26 | Cortex Visual Peripheral |
| Thal Pulv | 5080 | 10 | -30 | 2 | 6.78 | Diencephalon |
| No label | 264 | 34 | -46 | 0 | 5.4 | No description |
| No label | 352 | 32 | -50 | 6 | 4.07 | No description |
| No label | 624 | 26 | -46 | 20 | 4.54 | No description |
| No label | 224 | 4 | 8 | 26 | 4.33 | No description |

**Supplementary Table 4.** Moderation of *Path ab* effects by individual differences in explicit learning (FDR corrected $q < 0.05$).

| Region | Volume (mm3) | X | Y | Z | maxZ | Large-scale region or network |
|---|---|---|---|---|---|---|
| **Moderation of *path ab* effects by individual differences in explicit learning** | | | | | | |
| CA3 Hippocampus | 304 | -32 | -22 | -12 | 3.62 | Hippocampus |

**Supplementary References**